\documentclass[10pt,conference]{IEEEtran}
\usepackage[T1]{fontenc}
\usepackage[utf8]{inputenc}
\usepackage{times}
\usepackage{microtype}
\usepackage{url}
\usepackage{cite}
\usepackage{graphicx}
\usepackage{tikz}
\usetikzlibrary{arrows.meta,positioning,calc}
\usepackage{xcolor}
\usepackage{balance}
\usepackage{tabularx}
\usepackage{booktabs}

\definecolor{ctblue}{RGB}{41,91,134}
\definecolor{ctgreen}{RGB}{32,107,91}
\definecolor{ctgray}{RGB}{77,91,104}
\definecolor{ctgold}{RGB}{137,97,32}

\title{Cognitive Tours: Making Understanding a 
Deliverable in AI-Assisted Development}
\author{\IEEEauthorblockN{Margaret-Anne Storey}
\IEEEauthorblockA{Department of Computer Science\\University of Victoria, Canada\\mstorey@uvic.ca}}
\begin{document}
\maketitle
\pagestyle{plain}
\thispagestyle{plain}
\begin{abstract}
Generative AI and agentic development are accelerating software development, leading to concerns about technical debt. But two less visible forms of debt, cognitive debt and intent debt, are emerging as key risks to software health. \emph{Cognitive debt} is the absence or erosion of a team's shared understanding where no one can confidently explain how a system works or predict the impact of a change. \emph{Intent debt} is the absence of clearly articulated goals, constraints, and rationale that explain what the system is for and guide how it should evolve. 
This paper presents \emph{Cognitive Tours}, a tool designed to mitigate and repair cognitive and intent debt. 
Cognitive Tours is a web-based prototype for creating and following curated paths connecting user needs, design decisions, code, tests, and open questions. AI-assisted skills help audit existing waypoints and tours for possible problems or staleness and suggest candidates for new ones; humans decide what matters, verify relationships, and construct the narrative. This division of work deliberately preserves \emph{productive friction}: the sensemaking involved in selecting, explaining, and revising a tour, which is intended to foster and share understanding.  Our demonstration uses a tour of \emph{Cognitive Tours itself} to introduce the tool and illustrate its design decisions and interaction mechanisms. A proposed evaluation agenda concludes the paper.

\end{abstract}

\section{Introduction}

Generative AI and agentic development are accelerating software development, allowing individuals and teams to generate and modify code faster than ever before. But many developers and teams are sharing how they struggle to keep track of the work they are doing and answer questions about the reason for its design, the constraints it must satisfy and how the system can evolve to meet future requirements and changes~\cite{storey2026debt}.
Cognitive debt is held by a team, not only an individual: when the reasoning behind a change lives in one developer's head or one chat transcript, everyone else on the team, and that developer months later, inherits the gap. 

Although developers may have access to the code, agent interaction histories and  agents to explain how their systems work, developers struggle to maintain understanding and control of their system~\cite{tavantzis2026cognitivedebtgenaiera}. Writing and modifying code has traditionally provided opportunities for developers to develop their understanding through implementation decisions and problem solving. Delegating these activities to AI may remove some of those opportunities. Our prior research on troubleshooting~\cite{starr2026troubleshooting} and developer momentum~\cite{starr2026momentum} describes the cognitive work involved in building up working context to support progress on a task during development with AI.

Furthermore, the \emph{why} behind many of the decisions developers make is often not found in the source code (e.g., discarded ideas) and both humans and agents struggle to understand invisible but critical design decisions.

The \emph{triple debt} model of software health draws attention to two concerns beyond technical debt: \emph{cognitive debt}, the absence or erosion of the shared understanding needed to work with a system, and \emph{intent debt}, the loss or absence of explicit goals, constraints, and rationale \cite{storey2026debt}. A program may be technically sound and of high quality, but be difficult for its maintainers to explain or know how it can be safely evolved. Furthermore, a working and plausible implementation may not satisfy a requirement that was never adequately articulated or externalized.

\emph{Cognitive Tours} provides a mechanism for developers relying on AI and/or agents to hand craft curated journeys through the reasons, decisions, implementation, evidence, and uncertainties that help tell the story behind a feature or task. The intended user of these tours is not only a newcomer onboarding to unfamiliar code, but also the developer themselves who must later make changes to the system and recapture forgotten decisions and assumptions. 

Tours are human-authored by default, with optional AI drafting that requires human verification. The designer or developer must decide what matters to record and what to omit so that the important information in the tour is salient. 
AI supports the identification of new candidate and outdated waypoints and tours as the system evolves. The approach is designed to reduce the toil in creating and updating them, but not to automate them fully, which would suppress the understanding they build.  

This paper demonstrates a web-based prototype through a \emph{self-tour}: we use a tour of Cognitive Tours to explain the motivation and design of Cognitive Tours. 

Our contribution is a \emph{tool and interaction design proposition}, not a validated claim (yet) that tours can reduce or mitigate cognitive or intent debt. However, the tool has been used by the author during its development and provided cognitive support across sessions over six months and exposed a discrepancy between a design decision (captured as an ADR~\cite{nygard2011adr}) and a recent implementation choice. 

\section{Background and Tool Positioning}
Twenty years ago, my colleagues and I introduced the \emph{waypoint metaphor} for collaborative understanding and software navigation and orientation~\cite{storey2006tagsea}.

The \emph{Tours} Eclipse plugin combined slides and live IDE navigation for programmer-to-programmer presentations and tours~\cite{cheng2007tours}. 
In this paper, we revisit these design principles and 
adapt \emph{software tours} and \emph{waypoints} to today's era of AI-assisted development. 
Our earlier attempt saw barriers to adoption due to challenges maintaining waypoints and tours as the software changed over time.  Today, this toil can be mitigated through the use of AI and agent skills to help developers keep waypoints and tours up to date but preserve how creating them helps enrich understanding. The use of AI is limited to scaffolding while the developer is responsible for understanding.

Recent work has explored the potential value
of guided code exploration. 
Horvath~\cite{horvath2024} studies how developers’ thoughts, questions, design rationale, and development traces can be externalized as meta-information to support subsequent sensemaking. Cognitive Tours shares this emphasis on preserving developer reasoning, organizing it into human curated paths that connect intent, decisions, features, user needs, implementation, and tests.
Taylor and Clarke
studied Microsoft CodeTour, manually creating two annotated CodeTours, studying its potential for orientation and navigation~\cite{taylor2022tour}.
LACY combines AI-generated content
with expert-curated code tours, comprehension
questions, and other onboarding support~\cite{kara2026lacy}. Its evaluation shows promising evidence for expert-guided onboarding. 
CodeMap similarly explores
human--AI collaboration and uses AI to dynamically generate
hierarchical visualizations for exploring unfamiliar
codebases~\cite{gao2026codemap}.
Cognitive Tours differs from CodeTour, LACY and CodeMap in its application beyond code artifacts to design and intent artifacts, 
and its allocation of cognitive
work.

\begin{table}[t]
\caption{Tools that create an artifact for understanding, on four design dimensions}
\label{tab:dimensions}
\footnotesize
\setlength{\tabcolsep}{3pt}
\begin{tabularx}{\columnwidth}{@{}>{\raggedright\arraybackslash}p{1.45cm}
  >{\raggedright\arraybackslash}p{1.05cm}
  >{\raggedright\arraybackslash}p{1.15cm}
  >{\raggedright\arraybackslash}X
  >{\raggedright\arraybackslash}p{1.05cm}@{}}
\toprule
Tool & When & AI writes & What it records & For whom \\
\midrule
Spec Kit, Kiro & before code & much & requirements, plans & agent \\
Lattice & before, during & some & team standards, design decisions & agent \\
SpecStory & during & none (a log) & AI coding conversations & team, agent \\
CodeTour~\cite{taylor2022tour} & after & none & code walkthroughs & newcomers \\
LACY~\cite{kara2026lacy} & after & some & code tours, quizzes & newcomers \\
As-built architecture & after & all & architecture, workflows & people \\
Diagnostic Legibility & during & all & domain concepts, workflows & people \\
DeepWiki, Code Wiki & after & all & code documentation & anyone \\
Bram & during, after  & much; humans approve & proposals, approvals, sessions & team, agent \\
\midrule
\textbf{Cognitive Tours} & before, during, after & suggests; humans curate (AI drafts optional) & goals, needs, rationale, open questions, linked to code and tests & team, author \\
\bottomrule
\end{tabularx}
\end{table}

Another recent tool, Bram\footnote{\url{https://github.com/judell/bram}},
preserves agent-work sessions, human approvals,
and associated development history. 
Bram further provides AI-generated orientation
diagrams of a repository's architecture.
Cognitive Tours
complements Bram's approach by helping developers
construct and maintain an explicit account of
software intent and understanding.
Romeo et al.~\cite{romeo2024drift} also investigate drift between software design, implementation, and documentation, using links between UML diagrams and source code to identify inconsistencies as systems evolve. Cognitive Tours takes a complementary approach by connecting user needs, design rationale, code, and tests through curated navigation paths, with AI-assisted auditing to identify potentially stale or inconsistent explanations.

Balfroid et al. \cite{balfroid2024tours} explored LLM-generated code tours based on failing-test traces, finding that automated generation is feasible but can produce redundant and low-level explanations. 
Going further, AI-generated repository wikis such as DeepWiki\footnote{\url{https://deepwiki.com/}} and Google Code Wiki\footnote{\url{https://codewiki.google/}} document an entire codebase automatically, with no developer authoring.
Cognitive Tours instead uses AI to support discovery and maintenance while keeping developers responsible for the explanations and rationale.

\begin{figure*}[t]
    \centering
    \includegraphics[width=\textwidth]{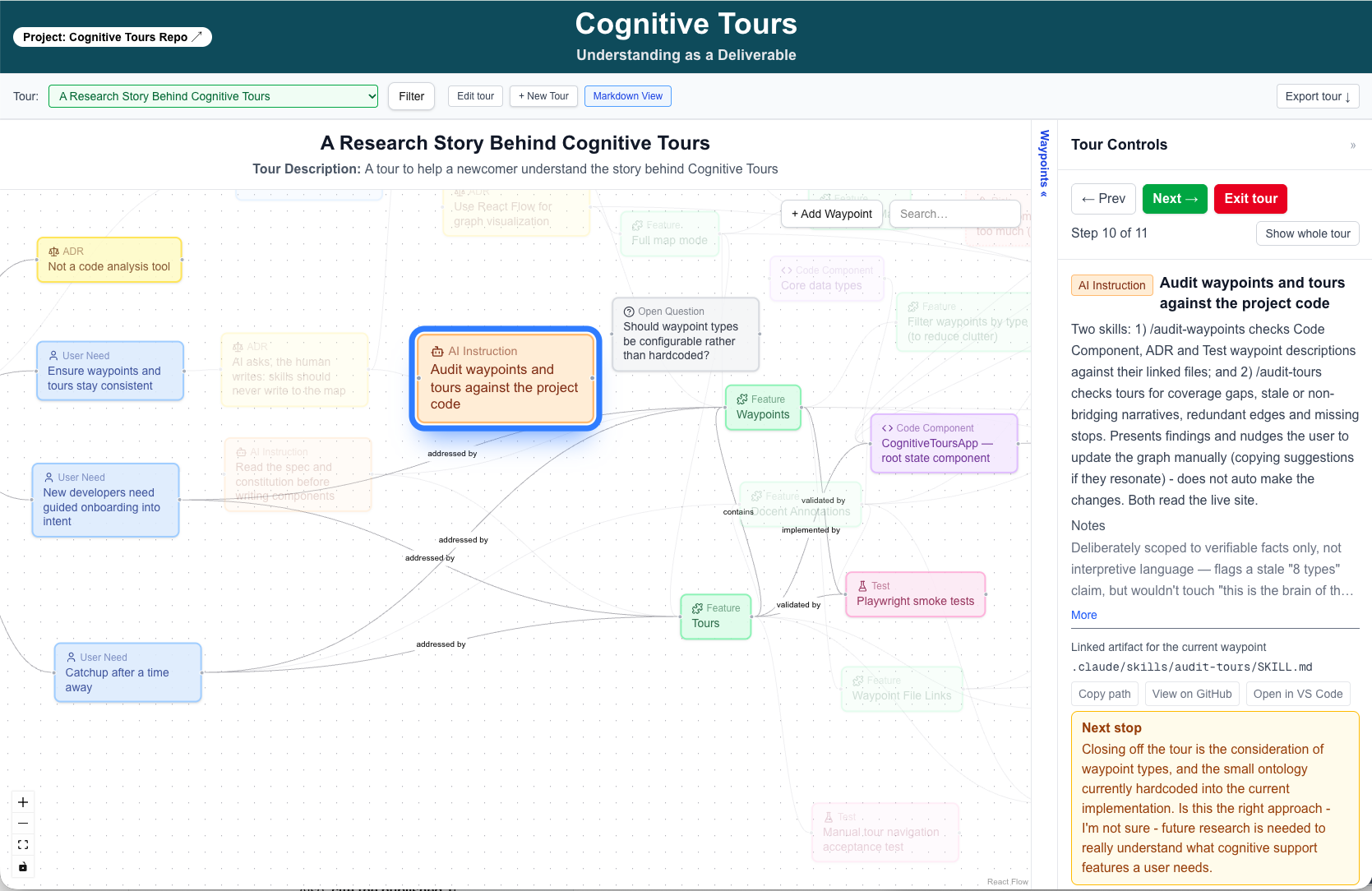}
    \caption{The Cognitive Tours interface showing a guided tour
    of the application itself. The waypoint graph connects
    user needs, design decisions, and implementation artifacts,
    while the commentary panel, which holds waypoint notes and docent annotations between tour stops, guides developers through  the reasoning behind the system.}
    \label{fig:cognitive-tours}
\end{figure*}

These tools differ along at least four design dimensions (see Table~\ref{tab:dimensions}): \emph{when} an account of the software is
constructed, \emph{how much} of it AI writes, \emph{what} it records, and
\emph{for whom}. They range from spec-driven tools that record intent before code
is generated (GitHub Spec Kit\footnote{\url{https://github.com/github/spec-kit}},
Kiro\footnote{\url{https://kiro.dev/}}), through tools that give the agent context
or record work as it happens
(Lattice\footnote{\url{https://github.com/techygarg/lattice}},
SpecStory\footnote{\url{https://specstory.com}}, Bram), to accounts generated
after the code exists
(as-built-architecture\footnote{\url{https://github.com/gregce/as-built-architecture}}, Habitat's Diagnostic Legibility\footnote{\url{https://github.com/Habitat-Thinking/ai-literacy-superpowers}},
DeepWiki, Code Wiki). 
Cognitive Tours records what otherwise stays in developers'
heads (goals, needs, rationale and open questions), links it to the code, and is
written by people for people: its skills audit and suggest, but people write the accounts for waypoints and tours.

All the tools in Table~\ref{tab:dimensions} make a representation available as a deliverable intended to support understanding: a specification, context file, log, tour, or guide. They differ in what is represented, who constructs it, and whose understanding it supports. Other interventions focus on developing understanding through questioning or reflection, without making a persistent account of the software their primary deliverable (e.g., Learning
Opportunities\footnote{\url{https://github.com/DrCatHicks/learning-opportunities}},
/grill-me\footnote{\url{https://www.aihero.dev/skills-grill-me}} and the Feature
Comprehension Score\footnote{\url{https://feature-comprehension-score.elliedocs.com/}}).

\section{Cognitive Tours}
\subsection{Main Features: Waypoints, Relationships, and Tours}
Cognitive Tours relies on three core concepts: a \emph{waypoint}, a \emph{relationship} or link between two waypoints, and a \emph{tour} of waypoints. A waypoint is a selected anchor or landmark that the developer considers important for understanding: a system goal, user need, feature, architecture or design decision, AI instruction, code component, test, risk, or unresolved question. Waypoints have both a description and optional commentary. A waypoint can link to an existing project artifact, such as a specification, ADR, source file, or test. Links open the artifact on GitHub or in the developer’s IDE.

Named (but intentionally untyped to be flexible) relationships explain why nodes are connected, for example a feature \emph{addresses} a user need, or a feature is \emph{implemented by} a component, or \emph{poses} an architectural question or decision. Waypoints are intentionally positioned on the graph by the user to help them preserve a \emph{cognitive map} of the system under development that persists across time.  A user composes a tour by selecting and ordering some of these waypoints to tell a coherent story for a particular feature or design decision. A tour normally traverses only some of the waypoint nodes in the underlying graph.  

Tours are shared artifacts: currently a map is reached through a single URL with no setup, each tour records its author and when it was last changed,
and each team works in its own map, so a tour one member writes becomes a path the rest of the team can follow and question (see Fig.~\ref{fig:cognitive-tours}). 

\emph{Docent annotations} provide human-curated commentary
between waypoints on a tour, explaining what matters,
what to notice, and how waypoints connect.
Creating these annotations encourages developers
to articulate rationale, including decisions and
alternatives that may not be recoverable from
the source code.

\begin{figure}[t]
    \centering
    \includegraphics[width=\columnwidth]
    {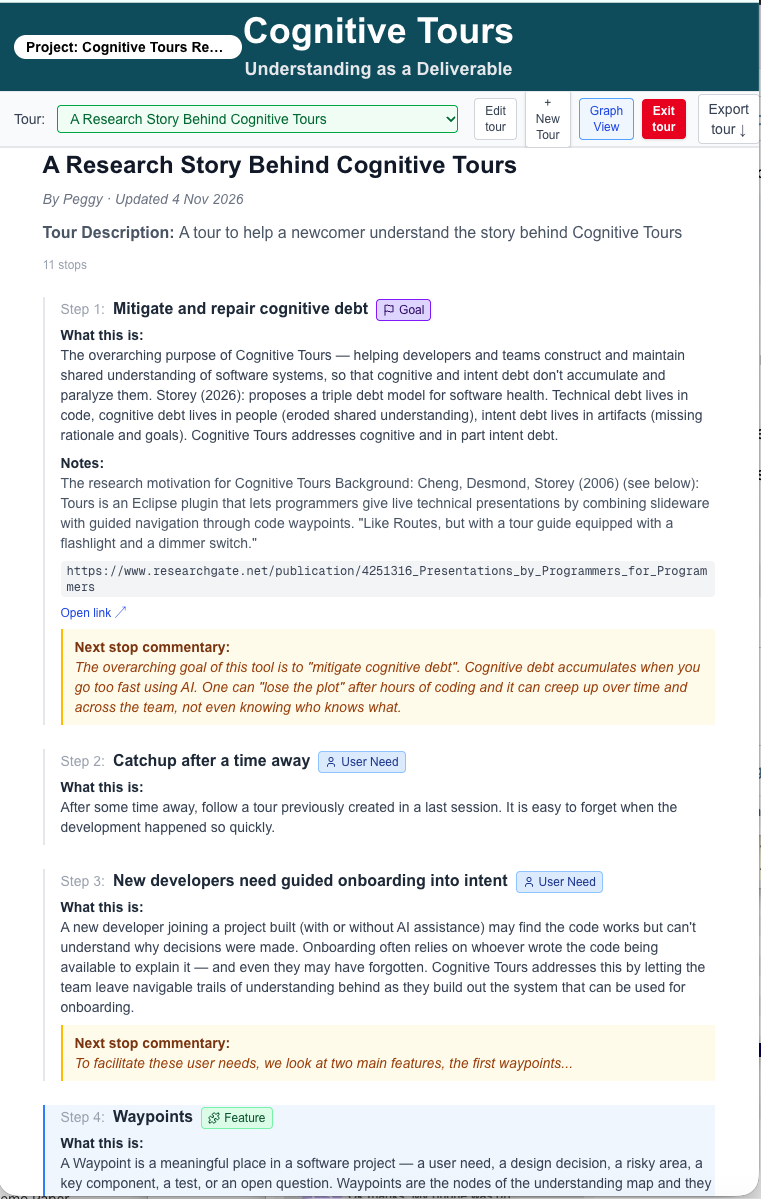}
    \caption{A tour can be shown (and saved) as a markdown view instead of a graphical view.}
    \label{fig:markdown}
\end{figure}

The browser interface is organized around a navigable waypoint graphical map and an explanatory, navigable tour view. The latter places and highlights the current stop in context and as mentioned above, presents human-curated commentary (docent annotations). The tours may be presented and navigated in the graphical view or as a markdown file (which can be exported, see Fig.~\ref{fig:markdown}). 

The tour authoring workflow allows the externalized information to evolve as the software or the team's understanding changes. Over time some waypoints and tours may become less relevant. Skills can recommend new waypoints and relationships, or updates to existing ones.

\subsection{AI-Supported Auditing, Human-Led Sensemaking}
Cognitive Tours uses AI-assisted skills to help maintain the waypoint graph and its curated routes. These skills audit individual waypoints and entire tours, flagging material that may need revisiting, and suggesting new waypoints or tours.  
A person still decides which suggestions are meaningful, verifies the underlying evidence, writes or revises the rationale and commentary, and determines the route through which someone else should encounter it. We deliberately retain the manual authoring and curation work, but use AI to reduce tedious discovery and maintenance work (observed blockers to our implementation from 20 years ago).

Four Claude Code skills accompany the tool:
1) \textbf{audit-waypoints}: finds waypoint descriptions that no longer match their files; 2) \textbf{audit-tours}: finds unrecorded recent work and stale tour narratives; 3) \textbf{suggest-waypoint}: ranks recent work worth recording, then asks the author to explain it; and 4) \textbf{suggest-tour}: proposes where new waypoints fit, and asks the author to write each bridge between them.

All four skills run in Claude Code inside the project's repository (see the demo). They read the live map through the tool's \texttt{/api/data}
endpoint, the git history since a given date, and the files that waypoints link to. They never write to the map. The two audit skills report findings
one at a time, such as a waypoint description that no longer matches its linked file, a tour stop narrative
that names a component that has since moved, or a recently changed file that appears in no tour; the two suggest skills rank recent work worth recording and supply only the fact fields (label, type, file path, edges, stop order), followed by questions for the author. The author decides which findings to act on and writes the descriptions and narratives (a skill drafts content only on request, after first prompting the author to write their own, so that they build their own understanding).

\section{Demonstration: A Tour of Cognitive Tours}

We demonstrate Cognitive Tours through a tour of the application itself, rather than an artificial example. A publicly accessible demonstration is available at \url{https://cognitive-tours-demo.vercel.app/}. To explore the tool, visitors can first select the ``A Research Story Behind Cognitive Tours'' tour from the application’s tour selector and navigate through its waypoints using the Next control.
The self-referential example (shown in the demo) serves two purposes:
it introduces the design and motivation of Cognitive
Tours while illustrating how a curated tour can
communicate the story behind a software system (see Fig.~\ref{fig:cognitive-tours}). This particular tour was created to onboard a new team member. Other tours are available for other purposes, such as documenting decisions made around feature designs. 

\textbf{Understanding why the system exists.}
The tour begins with the motivating problem:
AI-assisted development can accelerate software
production faster than developers can build and
share an understanding of what has been created.
The initial waypoints establish the user needs
and design goals.

\textbf{Following the design rationale.}
The tour then connects these goals to features
and design decisions. Rather than navigating an
unstructured collection of source files, users
follow an ordered sequence of waypoints accompanied
by contextual commentary (docent annotations). The underlying graph
makes relationships between goals, features,
decisions, and components visible.

\textbf{Connecting intent and implementation.}
The tour has waypoints representing intended user goals, design decisions, implementation
components and tests. The connections illustrate how a tour
supports movement between explanations of
why a feature exists and the technical artifacts
that realize it.

\textbf{Identifying open questions or gaps in understanding.}
Finally, the tour presents an unresolved question
about the system, such as whether waypoint types should be configurable. 
Open questions become part of the cognitive map indicating where further investigation is needed.

\section{Early Experience and Future Work}
Poor understanding is not only a risk for newcomers. Cognitive debt also accrues when we move fast. 
In one evening of AI-assisted work preparing the tool for a fall software engineering course\footnote{see \url{https://startup-studio.ai}}, I made more than 20 commits: a separate site and database for each of the student teams, project labels, links to each team's GitHub repository, and nightly backups.  Although I used Cognitive Tours to mitigate cognitive and intent debt through the tool development journey, during this particular coding session 
none of this was recorded by me in my project's own Cognitive Tours map. The next day, I ran a skill to compare recent commits against the map to surface what may be missing and should be recorded for future sessions. 
To my surprise, it ranked first a design decision that contradicted an outdated ADR. Had a collaborator joined the project the next
day, the map would have told them the opposite of
what the code did. I deferred answering, and wrote the waypoint and a new ADR some days later. 

The skill I executed (to help me repay the debt I had incurred) encouraged me to walk through what was built and update and add new waypoints and tours to capture this work; I recorded the new decision as its own ADR and marked the old one as deferred.
Figure~\ref{fig:suggest} shows an excerpt. The skill ranks decisions first, flagging any that contradict an existing ADR, and then asks questions rather than writing the waypoint; the description comes from my own answers. Although recording this information was tedious, even with AI assistance, it helped me rebuild context and regain momentum when I returned to development some weeks later~\cite{starr2026momentum}. Since this experience, I see the value in slowing down to record the decisions and choices that may matter in the future.

\begin{figure}[t]
\footnotesize
\begin{verbatim}
> /suggest-waypoint since 2026-10-20
Read the map: 59 waypoints, 73 edges,
5 tours; 28 commits since Oct 20.

Worth recording, in order:
1. ADR: One site and database per team
   <- contradicts ADR 007 (still Accepted)
2. Feature: Team sites
3. Feature: File links that work for everyone
4. Feature: Backups for every site
5. Risk: Team sites are open

> 1
- What was decided, in your words?
- Why was it needed? What would confuse
  someone without it?
- What else did you consider, and why not?

> A stopgap: a site and database per
  team, for my course and for demos.
  Why: time and ease of deployment.
  Later: one database, team-scoped rows.
\end{verbatim}
\caption{The \texttt{/suggest-waypoint} skill ranks unrecorded work and asks me about the top decision. My answer, given later and condensed here, became the waypoint's description.
}
\label{fig:suggest}
\end{figure}

We plan to compare human authored, AI-generated and partially AI-generated/audited
code explanations and tours to one another in terms of individual and team understanding.
We will assess orientation, behavioral
prediction, explanation quality, knowledge acquisition and knowledge
transfer, while considering the additional
effort required to create and maintain tours. We will also assess how cognitive tours can be used to foster a culture of understanding across a team, exploring how team members both curate and navigate each other's tours and show appreciation for the effort to do so. Finally, we will assess how writing a tour and the narratives for waypoints and tour stops enhances the author's own understanding, not only that of its readers.

\section{Discussion and Limitations}

The main aim behind Cognitive Tours is not to provide explanations of code. 
But rather it is to author deliberate developer accounts of  what matters, articulating the ``why'' behind decisions, reconciling intent with implementations, and preserving these accounts as the software evolves. 
We deliberately retain this
\emph{productive friction}, hypothesizing that
it supports the development of human understanding rather than merely the production of documentation.

This approach introduces costs and risks. Tours
and links may become stale, AI suggestions may
be inaccurate, and curated paths may omit
important steps. Human curation can
become an unnecessary burden. Whether the focus on human authorship leads to more accurate,
durable, and shared understanding than
AI-generated explanations remains future work.

\bibliographystyle{IEEEtran}
\bibliography{references}

\textbf{Acknowledgement of AI Assistance.} ChatGPT and Claude were used to suggest wording, improve clarity, and assist with manuscript revisions and formatting. Claude Code was used to assist with implementation of the prototype under the author's direction, and supported the writing of some of the demo tour narratives (as notes in the waypoints).

\end{document}